\documentclass[twocolumn,aps,citeautoscript,superscriptaddress,prb,10pt]{revtex4-2}
\usepackage{amsfonts,amsmath,latexsym,bm,wasysym}
\usepackage{graphicx,xcolor,blindtext}
\usepackage[utf8]{inputenc}
\usepackage{t1enc}
\usepackage{soul}

\newcommand*\chem[1]{\ensuremath{\mathrm{#1}}}
\newcommand{\dto}{\chem{Dy_2Ti_2O_7}}
\newcommand{\hto}{\chem{Ho_2Ti_2O_7}}
\newcommand{\ddto}{\chem{Dy_{2-x}Y_xTi_2O_7}}
\newcommand{\dhto}{\chem{Ho_{2-x}Y_xTi_2O_7}}

\begin{document} 

\title{Dipolar effects on the behavior of magnetically diluted spin-ice}

\author{M. Gorsd}
\affiliation{{Instituto de F\'{\i}sica de L\'{\i}quidos y Sistemas Biol\'ogicos (IFLYSIB), UNLP-CONICET, La Plata, Argentina}}
\affiliation{Departamento de F\'{\i}sica, Facultad de Ciencias Exactas, Universidad Nacional de La Plata, La Plata, Argentina}

\author{S. A. Grigera}
\affiliation{{Instituto de F\'{\i}sica de L\'{\i}quidos y Sistemas Biol\'ogicos (IFLYSIB), UNLP-CONICET, La Plata, Argentina}}
\affiliation{Departamento de F\'{\i}sica, Facultad de Ciencias Exactas, Universidad Nacional de La Plata, La Plata, Argentina}

\author{R. A. Borzi}
\affiliation{{Instituto de F\'{\i}sica de L\'{\i}quidos y Sistemas Biol\'ogicos (IFLYSIB), UNLP-CONICET, La Plata, Argentina}}
\affiliation{Departamento de F\'{\i}sica, Facultad de Ciencias Exactas, Universidad Nacional de La Plata, La Plata, Argentina}

\date{\today}

\begin{abstract} 
In this work, we explore the magnetic behavior of diluted spin-ice systems, where magnetic moments are randomly removed at various concentrations. We concentrate on features in which the effect of long range dipolar interactions (usually masked by self-screening in these systems) is made visible by dilution. Our initial focus is on the configurations reached after cooling to low temperatures at zero-field, sweeping the whole density range of impurities. We observe that the missing magnetic moments induce a certain type of local, magnetic charge order. Next, using Monte Carlo simulations, we examine the behavior of the magnetization under an applied magnetic field in the [111] crystallographic direction. 
The inclusion of dipolar interactions allows to account for the main features observed in previous  experimental results. Using the dumbbell model, where magnetic moments are represented as pairs of oppositely charged magnetic monopoles, we are able to understand the qualitative behavior of these curves as we increase doping. Additionally, we use this framework to calculate the critical fields corresponding to the phase transition observed in pure samples, and the characteristic fields appearing at very low doping.
\end{abstract}

\maketitle

\section{Introduction}

The low temperature states of geometrically frustrated materials present us with examples of magnetically disordered ground states in a context of perfect crystalline order.  Spin-ice materials \cite{bramwell_history_spin_ice_2020} are a good example of this phenomenon where, even in the absence of structural disorder, their uniquely constrained dynamics lead at low temperatures to disordered states with exceptionally long relaxation times, akin to glassy behavior \cite{snyder2001spin,samarakoon2022structural,hallen2022dynamical}. In these systems, the effects of added quenched structural disorder are far from predictable, as they can both create additional layers of disorder and longer relaxation times \cite{sen2015topological}, but also promote faster dynamics \cite{snyder2001spin} or lead to ordered states through the relief of frustration \cite{scharffe2015suppression}. 

\begin{figure}[b]
\includegraphics[width=0.7\columnwidth]{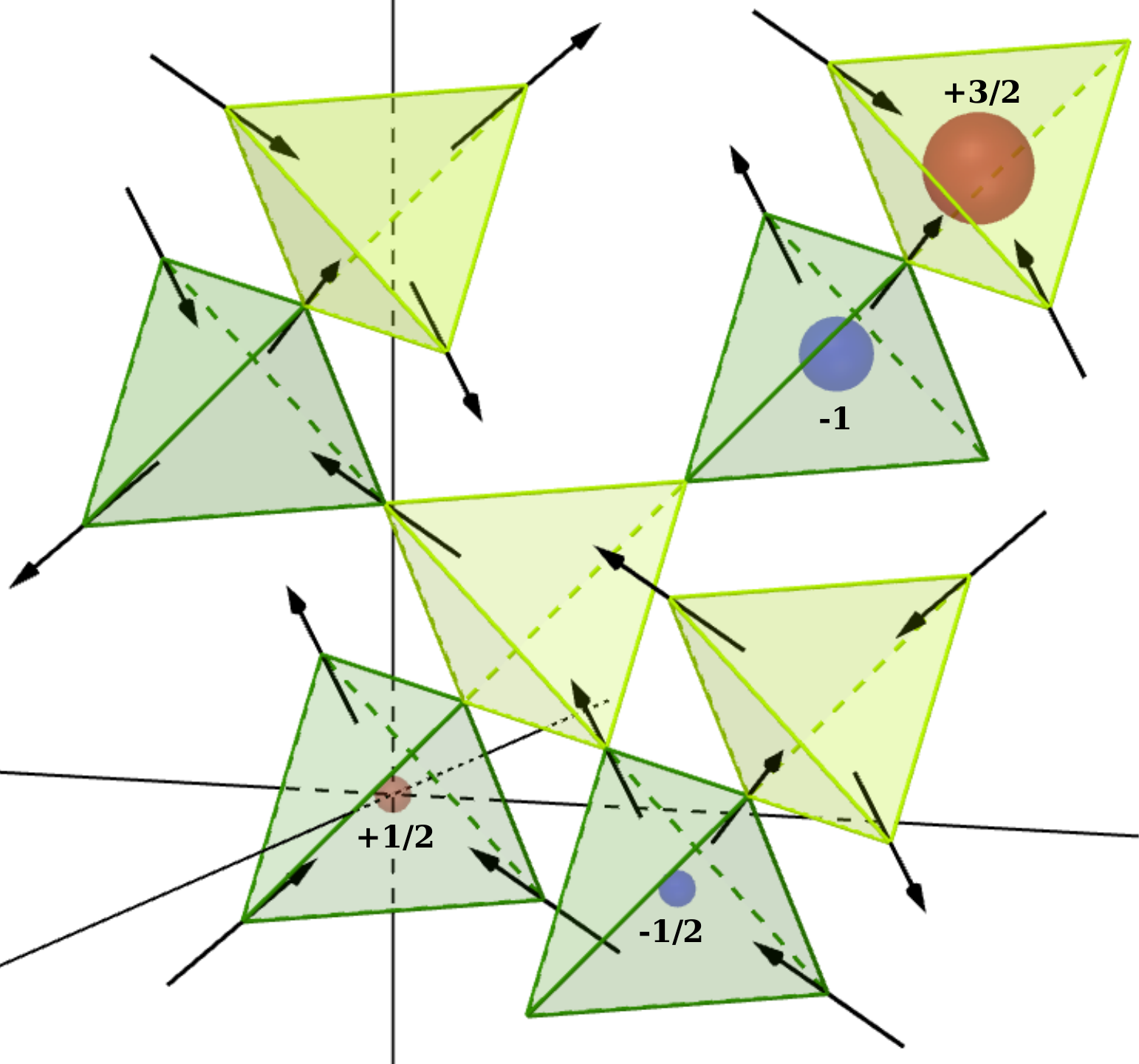}
\caption{Schematic view of a pyrochlore lattice, composed by ``up'' (light green) and ``down'' tetrahedra (dark green). We represent spins by arrows, and indicate monopoles using spheres of different color and size to indicate sign and absolute value of the topological magnetic charge. The locally neutral spin-ice state of pure samples (with two spins pointing in and two out from each tetrahedron), is modified by the introduction of diamagnetic atoms (missing arrows). This leads to irreducible fractional charges when the dilution in a tetrahedron corresponds to an odd number of magnetic moments.}
\label{fig:latt}
\end{figure}

In the context of the emergent Coulomb liquid characteristic of spin-ice materials \cite{castelnovo_2008}, magnetic dilution, in the form of doping with magnetically neutral sites, is of particular interest as it can be pictured as the introduction of fixed, fractional, magnetic charges. Experimentally this is achievable~\cite{snyder2004quantum,lin2014nonmonotonic} since the two canonical spin ice materials, \dto\ and \hto\ can be doped with non-magnetic Y$^{3+}$ ions. Both \ddto\ and \dhto\ form a solid solution in the whole range of doping, and the ionic radius of Y$^{3+}$ compared with Ho$^{3+}$ and Dy$^{3+}$ is sufficiently close that the lattice deformations caused by substitution are negligible \cite{AndPRX_2014,lin2014nonmonotonic}.  By combining specific heat measurements and simulations, Lin and collaborators \cite{lin2014nonmonotonic} showed that for \ddto\ this substitution preserves magnetic interactions and local anisotropy for $x<1.8$, and the system can be accurately modeled by a site-diluted dipolar spin model. Further specific heat measurements showed that a low level of non-magnetic impurities has little impact on the ground state, which still obeys spin-ice rules, while the residual entropy is eventually suppressed for $x>0.2$ \cite{scharffe2015suppression}.

The response to the application of external magnetic fields can give useful information about the states induced by magnetic dilution.  In spite of this, there are few in-field experimental~\cite{prabhakaran2011crystal,liu2015dy} and numerical~\cite{peretyatko2017interplay} studies of the magnetic properties of these systems with control of the magnetic field direction relative to the crystal axes. Experiments measuring magnetization with field along the crystallographic [111] direction \cite{liu2015dy} show a gradual softening of the features as the dilution is increased; in particular, at $T=0.5 {\rm K}$ the kagome-ice plateaux~\cite{sakakibara2003observation} is no longer present for $x \ge 0.4$.  This same case has been simulated numerically in the nearest-neighbor (NN) model~\cite{peretyatko2017interplay} where, instead, the two plateaux case of no dilution was replaced by a cascade of five (smaller) jumps of decreasing size as the dilution was increased. These jumps correspond to different possible distribution of vacant sites among neighboring tetrahedra.  This discrepancy between model and experiments is not surprising.  As already mentioned by the authors in \cite{peretyatko2017interplay}, the modeling of real materials necessarily involves the presence of dipolar interactions. This is not a small detail in the case of diamagnetic dilution: the projective equivalence between the dipolar and NN model is only valid within the strict 2 in - 2 out manifold~\cite{isakov_2005spin} which (as noted in Ref.~\cite{lin2014nonmonotonic}) ceases to exist even at the lowest temperatures and smallest dilutions.

Here we revisit the magnetic properties of diluted spin ice. Our theoretical and numerical study is based on the dipolar spin ice model, focusing on the specific effects caused by the inclusion of dipolar interactions on these doped materials. The quantitative and qualitative use of the dumbbell model serves to interpret our extensive Monte Carlo simulations in a way that appeals to our physics intuition, based on electrostatics. First we focus on the zero-field ground state of these systems, where we find that dilution induces local magnetic charge order, most likely linked with the reduction in the measured residual entropy ~\cite{lin2014nonmonotonic,scharffe2015suppression}. Next, we study the magnetic behavior under an applied magnetic field along [111]. Our Monte Carlo simulations of the magnetization as a function of the magnetic field are a good match to the experimental data. Further, combining standard solid-state physics techniques and the dumbbell model~\cite{brooks_artificial_2014,guruciaga2014monopole} we calculate the critical field corresponding to the polarization transition, and other characteristic fields for the diluted and undiluted materials. Finally, we construct a  qualitative framework using the dumbbell model that can successfully explain the smooth behavior of the simulated magnetization curves for intermediate  dilution levels.

\begin{figure*}[]
\includegraphics[width=2\columnwidth]{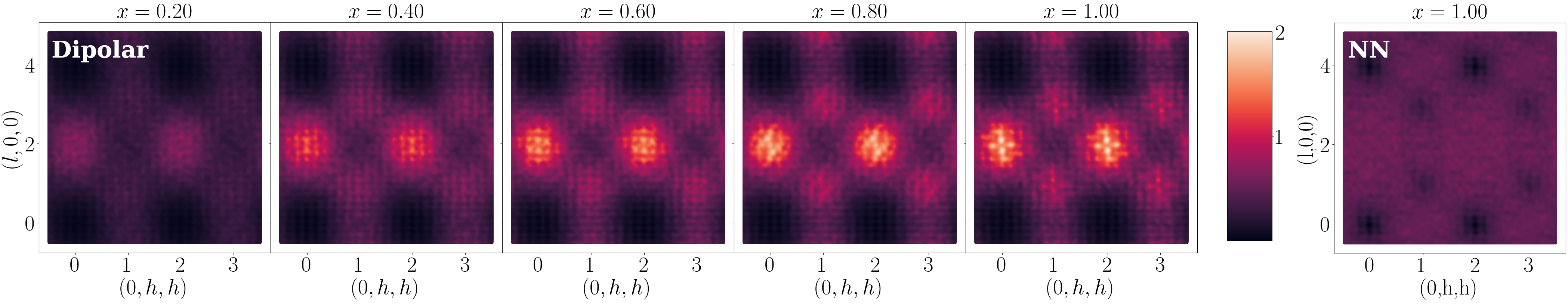}
\caption{Structure factor for the charge distribution at $0.3$~K and different degrees of magnetic dilution calculated for $L=6$ in the dipolar model. The figure shows from left to right  $x = 0.2,0.4,0.6,0.8$ and $1.0$.  The system is nearly symmetrical around $x=1$ so further dilutions are omitted. The last panel shows $S(q)$ at $x=1$ for the NN model.}
\label{fig:sq}
\end{figure*}

\section{Model and methods}

Spin-ice materials can be modeled as classical magnetic moments forming a pyrochlore lattice. These are Ising spins sitting in the corners of tetrahedra, and constrained to point along the local $\langle 111 \rangle$ directions (i.e., into or out of the tetrahedron where they belong).  The simplest model includes a nearest-neighbor ferromagnetic interaction, and is enough to describe the so-called \textit{ice rule} for the ground state. However, the large magnetic moments in these materials, $\approx 10 \mu_B$, require the presence of a dipolar term.  Additional terms, including up to third nearest neighbors, are necessary to account for some details in the thermodynamic properties of real materials such as \dto \ and \hto \cite{yavors2008dy, Borzi2016,henelius2016refrustration,samarakoon2020machine}. Here we will ignore these details in favor of a more general description, restricting the model to NN interactions and dipolar interactions only.

The exponentially degenerated ground state manifold of the pure spin-ice systems state is governed by the local, ice rule; it constrains any two spins to point in, and two to point out of each tetrahedron.  This rule can also be conceptualized as a discrete conservation law imposing a neutrally charged condition on the ground state, while excitations represent the creation of magnetic charges (single and double, with two possible signs in each case) that interact via a Coulomb law \cite{castelnovo_2008}. Magnetic moment dilution introduces quenched disorder in the lattice, and also the possibility of \textit{irreducible} fractional charges, that cannot be neutralized at any temperature. Even for a fixed number of missing spins there can be several possible topological charge values (see Fig.~\ref{fig:latt}), and several spin configurations for each charge.  For two missing moments the tetrahedron simply looses the possibility of a double charge excitation, but neutral and both positive and negative single charges are allowed.  For an odd number of missing moments there is a net irreducible charge in the tetrahedron: with one missing moment it can take the values $\pm 1/2$ and $\pm 3/2$, and with three missing moments it can alternate between $\pm 1/2$. At low temperatures, dilution implies that we need to rewrite the ice-rules in a more general way: within each tetrahedron, configurations that favor charges with the minimum absolute value are preferred.

For this work we performed Monte Carlo simulations in the pyrochlore lattice using the dipolar spin-ice model (DSM), defined by the Hamiltonian  
\begin{align}\label{eq:dsm}
    {\mathcal H}_{\mathrm{DSM}} =  & \ 
    J\sum_{\langle ij\rangle}{S}_i {S}_j + D\, r_{\mathrm{nn}}^3 \sideset{}{'}\sum_{i>j} \left[ \frac{\hat{s}_i \cdot{\hat{s}}_j}{|{\mathbf{r}}_{ij}|^3}  \right. \nonumber \\ & \left. 
    -\, \frac{3(\hat{s}_i \cdot {\mathbf{r}}_{ij}) (\hat{s}_j \cdot {\mathbf{r}}_{ij}) }{|{\mathbf{r}}_{ij}|^5} \right] S_i S_j \, .
\end{align}
Here the first sum over indexes that label sites on the pyrochlore lattice is restricted to NN, while the primed one starts from second nearest neighbors. $r_{\rm nn}$ is the nearest-neighbor distance, $r_{ij}$ is the distance between spins $i$ and $j$, $S_i$ can take values $\pm 1$, $D \equiv \mu_0 \mu^2 /(4\pi r_{nn}^3)$ and the $\hat{s}_i$ are unit vectors in the local $\langle 111 \rangle$ directions of the tetrahedra.  $J$ is an effective magnetic energy that includes NN superexchange ($J_{nn}$) and dipolar contributions ($J=[J_{nn}+5D]/3$); it is approximately 1.1~K for \dto, and 1.8~K for \hto; $D \approx 1.41$~K for both compounds, is the dipolar constant. Although we keep the discussion general, when showing our results as curves we used the parameters corresponding to \dto, which (due to its bigger $D/J$) is the one where dipolar effects are stronger. Ewald summations were used to take into account long-range dipolar interactions \cite{melko2004monte,borzi_2013} and simulated systems with $L\times L \times L$ conventional unit cells with periodic boundary conditions. Thermodynamic data as a function of field were collected using a single spin-flip Metropolis algorithm. Basic parameters, such as cell size, lattice parameter and magnitude of the magnetic moments for specific materials were taken from refs. \cite{yavors2008dy,Borzi2016}. 

In order to record data that would minimize the problems associated with slow dynamics~\cite{matsuhira2000low,snyder2004low}, we have used different cooling protocols. They were designed or chosen having in mind a double target: $i-$ they should offer the possibility to unmask the effect that diamagnetic impurities may have in the actual low temperature thermodynamics; $ii-$ at least in principle, they should be able to be reproduced by an experimentalist in the laboratory, so as to contrast our main results (that could be refined with the best spin ice Hamiltonian parameters on next iterations ~\cite{lin2014nonmonotonic,Borzi2016,samarakoon2020machine}) with experiments.

For the zero-field studies, we generated 40 random impurity distributions. Then, using a simple ZFC protocol, we cooled down from 1.1 K to the desired temperature without applying a magnetic field, splitting the interval into 30 points and waiting 1000 Monte Carlo Steps (MCS) at each one. From each distribution, we recorded 5 configurations with intervals of 1300 MCS. This process yielded 200 configurations per dilution $x$, over which the structure factor was averaged. In the case of magnetization vs field curves, we elaborated a protocol (FFC, field cooling at each field) that cools the spin system in a constant applied field (a \textit{field cooling}, FC) for each measured point.  The procedure is as follows: for each curve we simulate 200 points; for every point we do a FC from 1.4K to the desired temperature. This range was divided into 50 steps, with an equilibration of 2000 MCS at each one.  After the cool-down, we equilibrated for an additional 1000 MCS and used the next 2000 MCS to compute the average.  For each dilution level we constructed 20 different initial configurations starting from a full lattice and randomly selecting empty sites to the given density, $x$.

\section{Results}

\subsection{Low temperature manifold at zero field}

We first start examining the zero-field low temperature manifold of the diluted system.  As we mentioned before, dilution allows for the presence of half-integer charges in tetrahedra with an odd number of moments.  Calling $N$ the total number of magnetic sites, and $N_s$ the number of missing moments distributed at random in the system, the probability of having $n$ magnetic atoms remaining in a tetrahedron as a function of $y=x/2=N_s/N$ is given by a binomial distribution:
\begin{equation}
P(n) = \binom{4}{n} (1 - y)^n y^{4 - n}.
\end{equation}
Thus, the probability of having a fractional charge in a given tetrahedron ---the number density of charges at very low $T$---  becomes, $P_{\rm frac}=P(1)+P(3)=1/2\, (1-(x-1)^4)$. This probability is symmetrical around $x=1$ and very flat, with $P_{\rm frac} \approx 0.5,$ for $0.5 \le x \le 1.5$ , i.e., at these dilutions one expects around half of the tetrahedra to have fractional charges in the ground state.

Although the position of missing spins is randomly chosen, and the absolute value of the irreducible charges at low temperatures is constrained to the lowest possible value to ensure minimum energy, this does not determine the spatial distribution of the sign of these monopoles across the host diamond lattice. In order to study this, we calculated the structure factor $S(q)$ for the magnetic charge~\cite{slobinsky_field_2019,slobinsky_monopole_2021}, through the Fourier transform for the charge-charge correlation function:
\begin{equation}
    S(\boldsymbol{q})=\frac{2}{N}\sum_{\alpha\beta}\langle Q_{\alpha} Q_{\beta}\rangle\,e^{i\boldsymbol{q}\cdot \boldsymbol{r}_{\alpha \beta}},
\end{equation}
\noindent where $\langle \dots \rangle$ indicates thermal averaging, the greek indices $\alpha,\beta$ sweep the sites of the diamond lattice, $N/2$ is the number of tetrahedra, $Q_{\alpha}$ represents the topological charge at position $\boldsymbol{r}_{\alpha}$, and ${\boldsymbol{r}}_{\alpha\beta}$ is the distance between monopoles.

Figure \ref{fig:sq} shows  $S(q)$ for $L=6$ at different degrees of magnetic dilution after ZFC to $T=0.3
$~K. From left to right $x = 0.2,0.4,0.6,0.8$ and $1.0$ for the dipolar model; further dilution values are approximately symmetric around $x=1$, and are therefore omitted from the figure. These patterns indicate that there is no long range charge order for any $x$, and is notably similar to what is observed in monopole-liquid systems: a diffuse version of a zincblende crystal of alternating charges \cite{slobinsky2018}. We see that the intensity of the peaks grows with $x$, peaking at $x=1$, while their widths gradually narrows down; this indicates the growth of monopole crystallites with the density of irreducible monopole charges. On the other hand, $S(q)$ for $x=1$ and nearest neighbors (last panel of Fig.~\ref{fig:sq}) gives much shorter and wider peaks than its dipolar counterpart.

Naively, the marked difference between the two models may be disconcerting given the well known ability of the simple NN model to describe the low temperature properties of spin ice materials.  Indeed, the need for dipolar interactions was not initially evident~\cite{bramwell_history_spin_ice_2020}, despite the big dipolar magnetic moments present in the canonical materials. After noting the spurious effects induced by cutting off the dipolar interactions~\cite{siddharthan2001spin,den2000dipolar}, the idea of projective equivalence between interactions was developed~\cite{isakov_2005spin}. As in so many other cases, the dumbbell model~\cite{castelnovo_2008} can be used to explain this in a intuitive way. Within this framework, the main effect of dipolar interactions can be interpreted as adding Coulomb-like interactions between the existing monopole excitations. Given that the ground state of the pure system is locally neutral, the dipolar interaction should \textit{not} perturb the low temperature properties of the system in the temperature range where the Coulomb physics is a good approximation. However, and crucially for this work, diamagnetic dilution implies magnetic charges even at the lowest temperatures: it is thus to be expected a very different energy landscape on doping and, in particular, one that favors the formation of cristallites of magnetic monopoles of alternating charge. Following the dumbbell approximation we can thus understand the mechanisms behind Fig.~\ref{fig:sq}. At sufficiently low temperatures these charges are not dressed~\cite{sen2015topological} and their influence does not extend beyond the short range. This readily explains the approximate $x=1$ symmetry in the low temperature structure factor as it is a simple reflection of the symmetry in the abundance of tetrahedra with an odd number of moments, as mentioned before. The absence of interactions between monopoles implies less charge correlation and wider peaks for a given density $x$, as observed for the NN model.

The average cluster size we measured from low temperature snapshots is surprisingly small; it ranges from near $2$ for $x=0.1$ to above $5$ magnetic monopoles for $x=1$. This may be partially understood if we consider the long range character of the interactions combined with the competing nature of the Coulomb interaction in charges with alternating sign; to this, we still have to add the disorder necessarily present in the system. As expected, the distributions are quite wide. Using the square root of the variance to characterize it we obtained for a system with $L=6$, a width of $\approx 2$ monopoles for $x=0.1$ and near $15$ monopoles for $x=1$ (we found a cluster as big as 384 magnetic charges in one of its realizations). In terms of shape, the clusters tend to be quite filamentary (each monopole tends to have a reduced number of neighbors with the opposite charge). One should bear in mind that the maximum charge density attainable ($P_{\rm frac}(x=1)=0.5$) is slightly above the site percolation threshold for the diamond lattice ($p_c \approx 0.43$~\cite{gaunt1983series}); further details concerning charge distribution will be published elsewhere. In the NN model, both the average cluster size for $x=1$, and the width of the distribution are between $2$ and $3$ topological charges: even at the maximum monopole density the cluster sizes are no bigger the ones for the dipolar model at $x=0.1$. This increase in monopole crystallite size due to the added dipolar interactions could explain, at least in part, the material-specific reduction in residual entropy with respect to the estimations made extending Pauling's procedure~\cite{ke2007nonmonotonic}, experimentally determined in spin ice materials~\cite{lin2014nonmonotonic,scharffe2015suppression}.

\begin{figure}[b]
\includegraphics[width=\columnwidth]{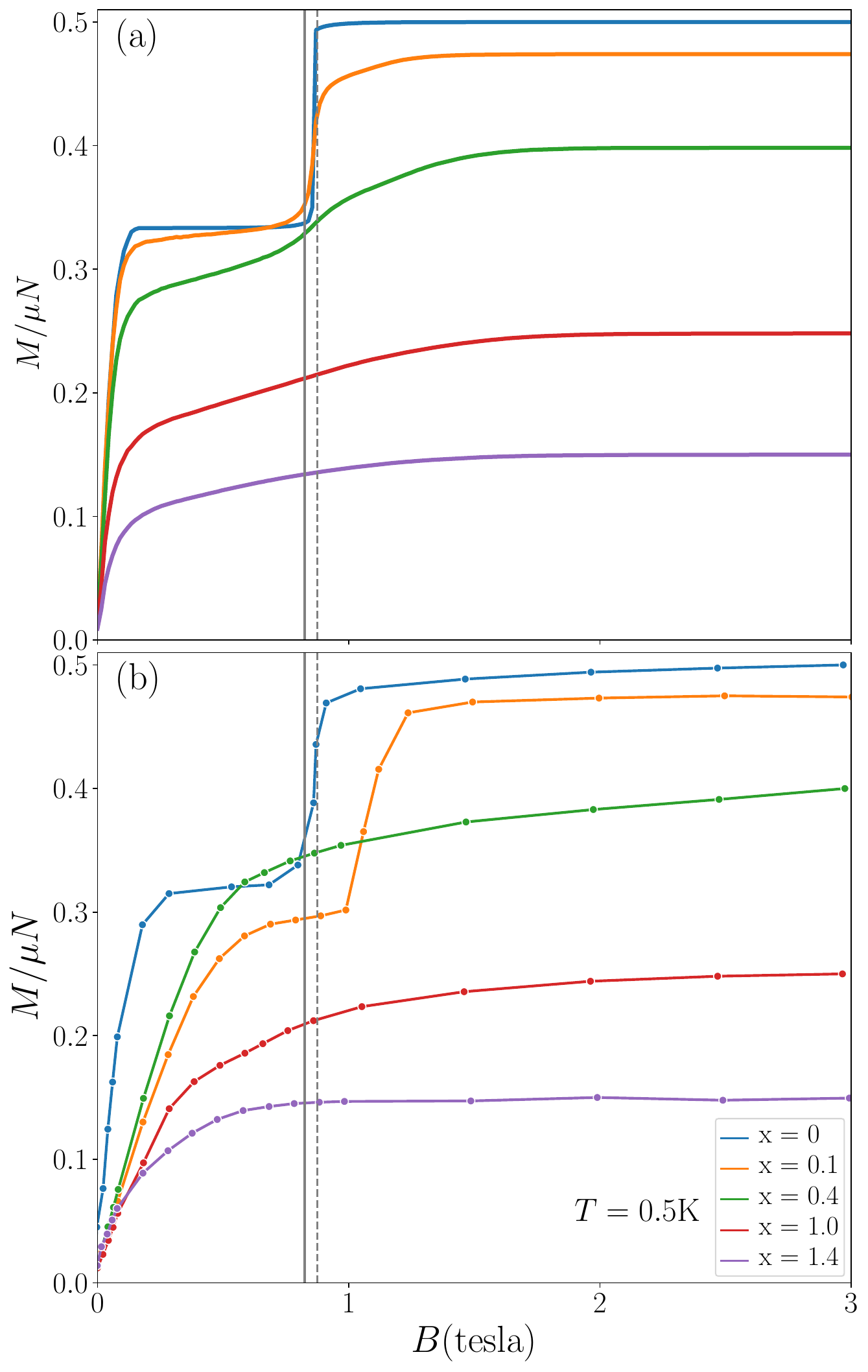}

\caption{ Comparison between the magnetization measured in \ddto\ samples by Liu,  et al \cite{liu2015dy} (measured with increasing field), normalized by the expected saturation value, and simulations using the dipolar model (measured with the FFC protocol), both at $500$~mK. The characteristic features of the pristine sample, a kagome-ice plateau and a sharp transition into full polarization, are softened by the dilution of magnetic moments. The dashed vertical line, ($B=$0.875\,T) corresponds to the experimental value of the critical field~\cite{aoki2004magnetocaloric}, and the solid one ($B=$ 0.825\,T) to the simulated value in thermal equilibrium~\cite{castelnovo_2008}. Both were obtained by extrapolation to zero temperature in the undiluted case.}
\label{fig:m1}
\end{figure}

\subsection{Magnetization with field along [111]}

We now turn to the response of the system as a magnetic field is applied precisely parallel to [111].  Along this crystallographic direction it is useful to picture the pyrochlore lattice as a stacking of triangular and kagome planes~\cite{sakakibara2003observation}.  In the pristine case, with no missing magnetic moments, the magnetization along [111] is characterized by two low temperature plateaux; as the field is increased there is a linear growth of the magnetization, driven by the moments in the triangular planes which eventually polarize along [111] without breaking the ice rules.  This is the kagome-ice phase (KI)~\cite{matsuhira_field_2002,higashinaka2003anisotropic}.  As the field is increased the magnetization remains in a plateau until at a critical field primarily (though not exclusively, as we will see) determined by the interaction $J$, it jumps through a first order phase transition into the fully polarized state~\cite{sakakibara2003observation}. In the language of charges, this jump was interpreted as the sudden condensation of a fully ordered single monopole crystal out of the locally neutral KI manifold~\cite{castelnovo_2008}.

Since within the NN model there is no monopole attraction, the sudden condensation of a crystal is replaced in this context by a crossover, with a smooth magnetisation curve. Notably, dilution drastically changes the shape of this curve.  As reported in ref. \onlinecite{peretyatko2017interplay}, at low temperatures it breaks up into four plateaux and intermediate jumps that result from tetrahedra with different number of vacant moments. This is in contrast to the experimental results on doped spin-ice materials, such as \ddto\ \cite{liu2015dy} (see panel a) of Fig.~\ref{fig:m1}). In the experiments, rather than introducing new plateaux, the dilution softens the existing features and washes out all traces of the KI plateau by $x=0.4$. We will argue now that this discrepancy is an expected consequence of the strong dipolar interactions beyond nearest neighbors.

Indeed, the long-range nature of dipolar interactions invalidates the local tetrahedra picture that explains the steps in the nearest neighbor model. Again, the magnetization behavior is best understood in the monopole scenario: in doped samples, all jumps are associated with a variation on the local charge distribution~\cite{peretyatko2017interplay}; hence, monopole attraction should play an important role. For moderate $x$ and low temperatures, the random distribution of interacting irreducible charges resulting from dilution creates a vast array of possible environments at each site. Since these environments are likely to be unique, the critical field associated with each one will also vary. This results in a continuous growth of magnetization as a function of field. A similar reasoning will help us understand some simple cases in a quantitative way (see Sec.~\ref{sec_crit-fields}).

Figure \ref{fig:m1} shows a comparison between our simulations using the dipolar model with the FFC protocol and the experimental results for \ddto\ of Liu et al.~\cite{liu2015dy} (measured with increasing field), both performed at $0.5$~K.  A direct quantitative comparison between our simulations and the experiment is not possible since the measurement protocols are different, and our model is missing the second and two third nearest neighbor interactions necessary for a quantitative modeling of \dto \cite{yavors2008dy,henelius2016refrustration,Borzi2016,samarakoon2020machine}. Nonetheless, the dipolar model captures qualitatively the transformation in the magnetization curve in diluted samples and shows the same tendency towards a softening of jump and plateaux observed in the experiments.

\begin{figure}[bh]
\includegraphics[width=\columnwidth]{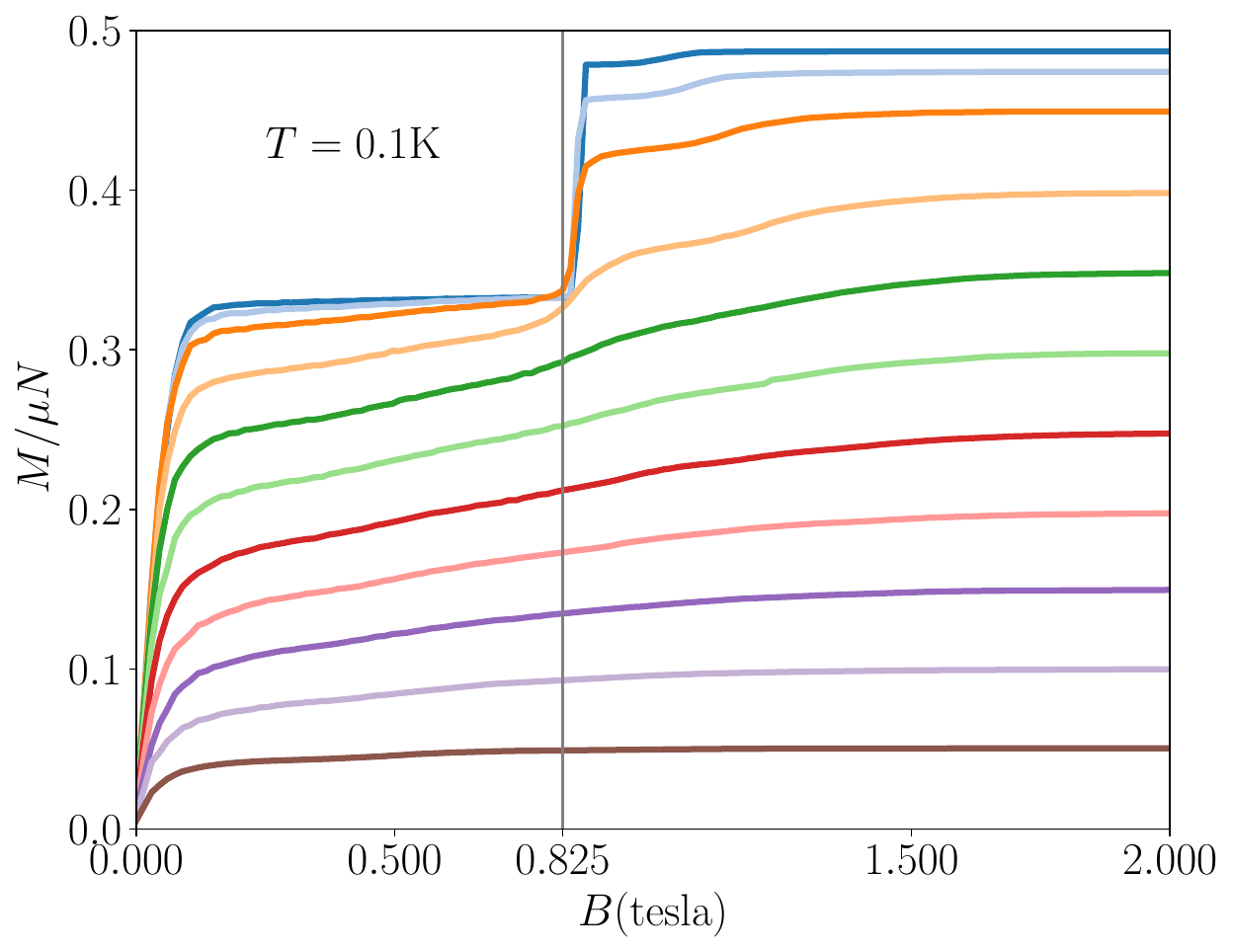}
\caption{Magnetization curves simulated for \ddto \ with field along [111] at low temperature, $T=0.1$~K,  for a wide range of dilution densities. From top to bottom $x= 0.05, 0.1,0.2$ and up to $x=1.8$ at 0.2 intervals. The vertical line corresponds to the critical simulated field for the undiluted case~\cite{castelnovo_2008}. Aside from the smoothing of the curves for increasing $x$, we notice a feature above the critical field that repeats for samples with very low dilutions, $x \lesssim 0.1$; it is associated with environments where a single spin in a triangular plane is missing (see Fig.~\ref{fig:defects}, top panel).
}
\label{fig:M2}
\end{figure}

A set of magnetization curves simulated at a lower temperature, $T=0.1 {\rm K}$ (FFC protocol), and for a wide range of dilution densities is shown in Fig. \ref{fig:M2}. While at low temperatures and low dilution levels the KI features and a polarization high-field transition are resolvable, the softening of features due to dilution at higher $x$ is still present. Remarkably, below $x=0.2$ there is a clear jump above the critical field where the monopole crystal condenses. The fact that these curves correspond to very low concentration of diamagnetic impurities suggest that it may be related with an environmental where a single missing spin (producing two irreducible charges) is relatively distant from other diamagnetic impurities. However, the characteristic field $B_{\rm{ch}}$ of this feature does not coincide with any of the jumps predicted by the NN model; the fact that at very low dilutions the disorder present is minimal will give us the possibility to predict the position of this jump using the dumbbell approximation. Interestingly, we will recall that the FFC protocol we use on the simulations could be repeated on a real experiment, allowing for a closer experimental contrast.

\subsection{Calculation of the critical fields for the dipolar model} \label{sec_crit-fields}

We will now use the dumbbell model to estimate some characteristic fields in the model with dipolar interactions in pure and diluted systems. This framework significantly reduces the complexity of calculations by focusing on a small number of magnetic charges, rather than accounting for the full intricacies of multiple interacting dipoles. Additionally, it enables the application of methods traditionally used in electrostatics.

In the dumbbell model \cite{castelnovo_2008} magnetic charges follow a Coulomb law, $U(r_{\rm ij})=\mu_0/ 4\pi~Q_{\rm i}Q_{\rm j}/r_{\rm ij}$, 
where $r_{\rm ij}$ is the distance between charges i and j.  Charges are measured in units of $2\mu/a_{\rm d}$, with $\mu$ the magnetic moment and $a_{\rm d}$ the lattice constant for the dual diamond lattice.  For \dto, $\mu = 10 \mu_B$, and  $a_{\rm d} = 4.33$ \AA.  Creating a charge in these systems carries an energy cost $U_Q=1/2 v_0 ~Q^2$, where 
\begin{equation}
    v_0 = \frac{a_{\rm d}^2 k_B}{\mu^2}\left(\frac{J_{nn}}{3}+\frac{4}{3}[1+\sqrt{2/3}]D\right).
\end{equation}
$J_{nn}=-3.72~{\rm K}$ ($-1.56~{\rm K}$) for \dto\ (\hto),  and $D=1.41 {\rm K}$ for both.

In order to illustrate the procedure, we begin by calculating the critical polarization field for \textit{undiluted} samples, extending to applied fields a method that has been previously used at zero field~\cite{brooks_artificial_2014,guruciaga2014monopole}.  The low field state corresponds to a neutrally charged state, while the high field one corresponds to a zincblende structure with a single positive or negative charge at every site of the diamond lattice. The change between the low and high field states corresponds to collectively flipping the minority basal spin in each tetrahedron; this corresponds to an energy difference per up-tetrahedron given by
\begin{equation}
    \Delta E = v_0Q^2 - \alpha\frac{\mu_0Q^2}{4\pi a_{\rm d}}-\frac{2}{3}\mu B
\end{equation}
The first term accounts for the energy cost for the charges created at the transition, the second corresponds to the interaction between them, with $\alpha =1.638$ the Madelung constant, and the third is the difference in Zeeman energy. Setting $\Delta E = 0$ we obtain 
\begin{equation}
    B_{c}(x=0)=\frac{3Q^2}{2\mu} \left( v_0 - \frac{\alpha \mu_0}{4\pi a_{\rm d}} \right) ;
\end{equation} 
Unlike in the NN model, and as a consequence of the long range dipolar interactions, this field depends on $J_{nn}$ and $D$ in a combination different from the effective exchange $J$.
This gives  $B_{c}(x=0) \approx 0.8244\,{\rm T}$ for \dto \,
in very good agreement with the value 0.825\,T numerically determined in ref. \onlinecite{castelnovo_2008} for the dipolar model. In the case of \hto\, $B_{c}(x=0) \approx 1.4675\,{\rm T}$.

If instead of collective moves we think on the field driving the flip of a \textit{single} minority basal spin in a locally neutral KI configuration (or the converse, in one of the single monopoles that populate the monopole crystal) we would be estimating the maximum (minimum) field at which the neutral vacuum (single monopole crystal) is metastable. The value we obtain for these ``spinodal'' fields are $B_{\rm{max}}=1.26\,\rm{T}$ and $B_{\rm{min}}=0.38\,\rm{T}$ for pure \dto\ and $B_{\rm{max}}=1.90\,\rm{T}$ and $B_{\rm{min}}=1.03\,\rm{T}$ for pure \hto.

\begin{figure}[bh]
\includegraphics[width=\columnwidth]{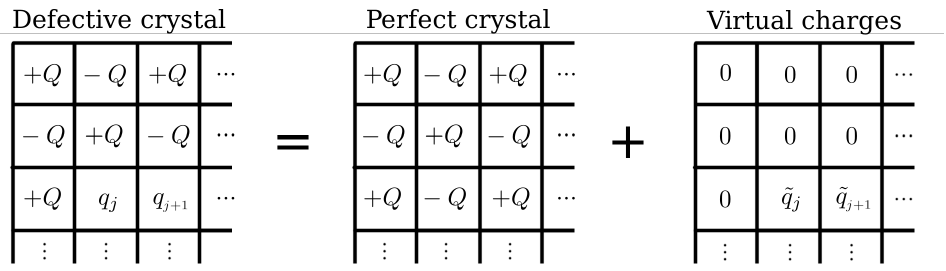}
\caption{ Schematic picture showing how a method of virtual charges can be used to calculate the energy of a defective crystal. Using the superposition principle, the potential is written as a sum of the potential of a perfect crystal plus the potential resulting from a finite set of virtual charges that take into account the defective sites.}
\label{fig:virtual}
\end{figure}

The dumbbell model also gives us the tools to calculate the characteristic fields of a system with missing moments. To the best of our knowledge, this is the first use of this tool in an impure system; a comparison with Ewald summations will show that the procedure provides quantitatively sound results.

For fields $B>B_c$ instead of a few charges in a vacuum, or a set of identical charges forming a crystal (which are the contexts in which this modeling is easier to apply) there would be an extensive number of charges of different magnitudes present in the system. However, using the  superposition principle, we can simplify this situation in the limit of small doping. There, one can express the potential of a defective crystal as the sum of the potential of a perfect crystal plus the potential of a small set of virtual charges that are non-zero only at the defective sites. We have schematized this idea in Fig. \ref{fig:virtual}.  In the defective crystal the charges take, almost everywhere, the values $(-1)^{i+1}Q$. If we refer to the charges at the defective sites as $q_i$, then, to obtain the defective crystal configuration from the perfect one, we must add to it a virtual charge $\tilde{q}_i=(-1)^i Q + q_i$ at each site where there is a defect. In this way, the potential at the site $i$ of the defective crystal can be written as the sum 
\begin{equation}
V_i = \frac{\mu_0}{4\pi} \sum_{j \neq i} \frac{(-1)^{j+1}Q}{r_{ij}} 
+ \frac{\mu_0}{4\pi} \sum_{j \neq i} \frac{\tilde{q}_j}{r_{ij}}.    
\end{equation}
Since the first term corresponds to a perfect crystal we can use Madelung's calculation to obtain
\begin{equation}
    V_i = (-1)^i \alpha \frac{\mu_0 Q}{4\pi a_d} + \frac{\mu_0}{4\pi} \sum_{j \neq i} \frac{\tilde{q}_j}{r_{ij}}.
\end{equation}
The energy of the system can then be calculated as the sum $U=1/2\sum_i Q_iV_i$.
\begin{equation}
    U
= - \frac{1}{2} N \alpha \frac{\mu_0 Q^2}{4\pi a_d} 
+ \alpha \frac{\mu_0 Q}{4\pi a_d} \sum_j (-1)^j \tilde{q}_j 
+ \frac{1}{2} \frac{\mu_0}{4\pi} \sum_{j \neq \beta} \frac{\tilde{q}_j \tilde{q}_\beta}{r_{\beta j}}.
\end{equation}
Here the first term corresponds to the energy of the perfect crystal, the second is the sum of the interactions of the virtual charges with the infinite crystal and the third the interaction between the virtual charges themselves.  This is a very useful expression when dealing with a reduced number of defects. 
\begin{figure}
\includegraphics[width=\columnwidth]{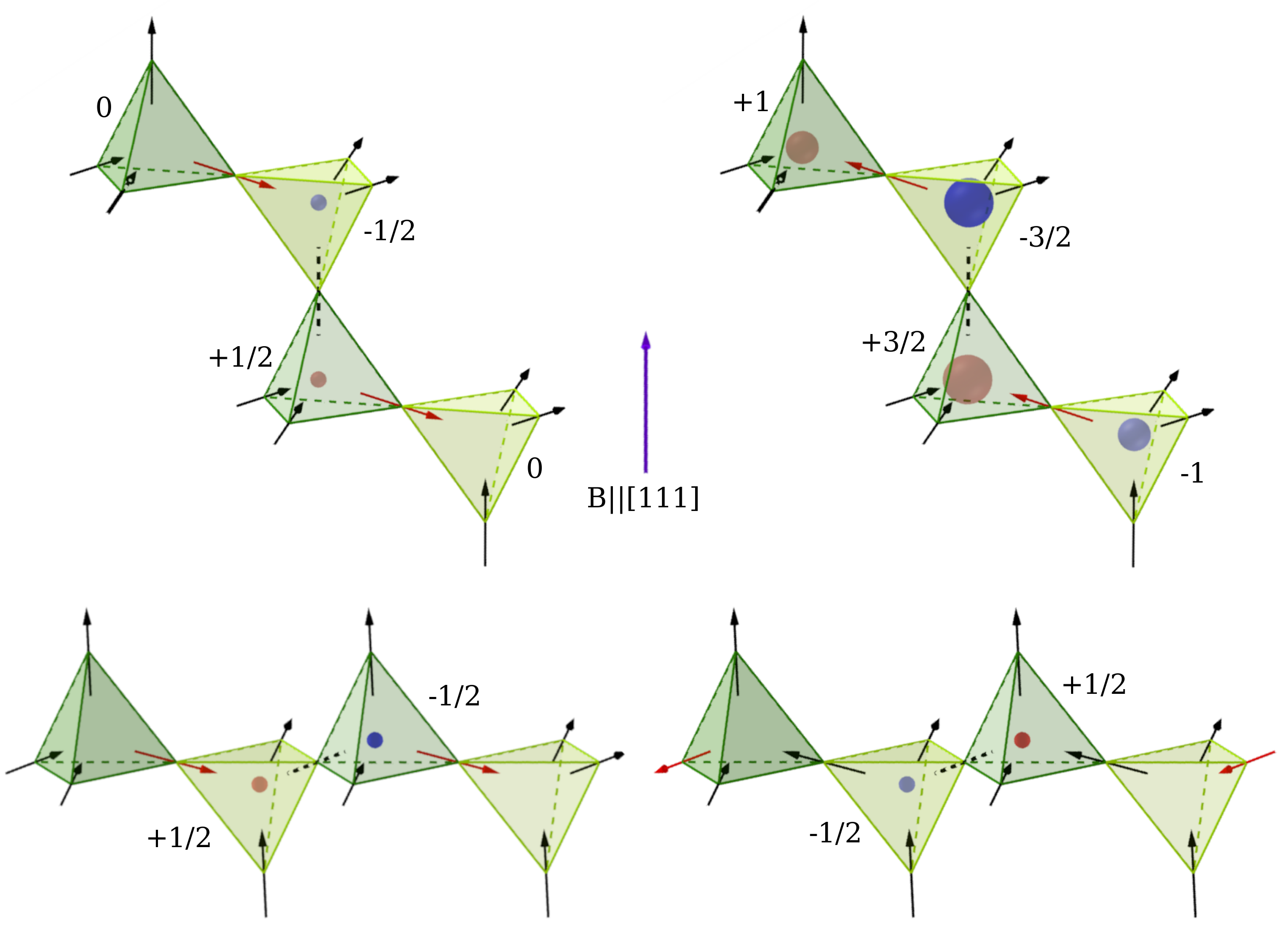}
\caption{ Schematic picture showing one defect in a triangular layer (upper panel) and a kagome layer (lower panel).  The missing moment in each case is marked with a dashed arrow.  The left side shows the low-field configuration and the right side the high field one.}
\label{fig:defects}
\end{figure}

We can readily study the case of single site doping, which is a good approximation to the situation for $x \ll 1$. We begin by looking at the case with one defect on the triangular plane; within the framework of NN interactions the apical spin favors the flip of the minority basal spin, for what we expect a characteristic flipping field $B_{\rm{ch}} > B_c(x=0)$ when it is missing. We will now extend these ideas to the case where dipolar interactions are present, using the dumbbell model.  Removing a triangular site results in the creation of a pair of $\pm 1/2$ charges in the low field case (see Fig.~\ref{fig:defects} upper left side) that become two $3/2$ charges in the high field side (see Fig.~\ref{fig:defects} upper right side).  There are two different scenarios depending on whether the resulting critical field would be below or above $B_c(x=0)$.  In the former the transition would happen on a neutral background (represented in the upper left side of Fig.~\ref{fig:defects}) while in the latter the background would be a zincblende crystal with one charge per diamond site. If we assume the first case we find that the difference in energy is
\begin{equation}
\Delta E = 3 v_0Q^2 +\frac{\mu_0Q^2}{4\pi a_{\rm d}}\,\delta
-\frac{4}{3}\mu B,
\end{equation}
where $\delta = 3\sqrt{3/8} - 5 - ((7/3)^2 + (2/3)\sec^2(\pi/6))^{-1/2}$. This gives a characteristic field $B_{\rm{ch}} \approx 1.7\,{\rm T}$ for \dto\ and $B_{\rm{ch}} \approx 2.66\,{\rm T}$ for \hto, inconsistent with the initial assumption.  

In the case of a monopole crystal background the factor $\delta$ in the second term changes to $\delta' = 1 - 4\alpha - \sqrt{3/8} - ((7/3)^2 + (2/3)\sec^2(\pi/6))^{-1/2}$ and results in a characteristic field $B_{\rm{ch}} \approx 0.945\,{\rm T}$ for \dto\ and $B_{\rm{ch}} \approx 1.91\,{\rm T}$ for \hto. The value obtained for \dto\ is compatible with the jump observed for $x \ll 1$ at fields above the critical field $B_c(x=0)$ in Fig.~\ref{fig:M2}. Also, we have numerically checked this result performing a simulation for the case of a lattice where a single triangular site was removed. For \dto\ we find that $B_{\rm{ch}}^{\rm{num}} \approx 0.95\,\rm{T}$, while for \hto\, $B_{\rm{ch}}^{\rm{num}} \approx 1.92\,\rm{T}$.

It is interesting to contrast these values for the characteristic field with those coming from the nearest neighbors model. For \dto\ (where the biggest differences are to be expected) in the same situation we would expect the much bigger field of $1.48$~T, a factor of $\approx 1.5$ due to the breaking of projective equivalence.  Within the dipolar model, the Coulombian interaction between the crystal that has been already stabilized by the field and the charge at the impurity sites is what favors the spin flip at a much lower field. The situation of very low dilution then offers a very direct way to observe tangible effects of the dipolar interactions in the relatively simple experiment of measuring the magnetization at low temperatures.

Of course, in a situation of low dilution it would be even more probable to remove a spin from kagome than from triangular planes. It is fair to ask for the magnetization jump associated with this situation: we analyze now that. If we remove a single site in the kagome plane we have two possible low energy configurations (small charges of opposite sign in NN tetrahedra) shown in the bottom panel of Fig. \ref{fig:defects}. These have the same number of charges, but different magnetization.  This is a subtle effect: although the configurations seem equivalent, in one case the moments opposed to the field are in the center of the configuration while in the other they are at the border.  In the infinite limit, or in the case of periodic boundary conditions like our simulations, field will select the configuration on the right~\footnote{One alternative way to see this is to consider a string of flipped spins connecting one charge in Fig.~\ref{fig:defects} bottom left to the other (the ``open loops'' of ref.~\cite{lin2014nonmonotonic}). Another one is to see it a flipping the \textit{ghost spin} of Ref.~\cite{sen2015topological}. It is easy to see that on introducing such a string we would be changing the sign of the two charges to accommodate this configuration into that of Fig.~\ref{fig:defects} bottom right, and also increasing the magnetization in the direction of a [111] field.}. The process does not imply the creation of new charges. More importantly,  the presence of any finite field along [111] always favors the same configuration: its characteristic field is $B=0$, and is only observed (both in the NN and dipolar model) as the initial increase of the magnetization.

The quantitative analysis in terms of charges becomes increasingly laborious as the number of defects in the sample grows. Nonetheless, it provides a valuable qualitative understanding of the magnetization curve. As the number of virtual charges is increased their abundance creates a wide variety of possible environments for each site, making it increasingly probable that each environment would be unique. Consequently, the characteristic field associated with each site becomes also unique, leading to a smooth increase in magnetization as monopoles are progressively generated.

\subsection{Symmetry about $x=1$, and different vacua}

Using the monopole picture, we have explained in a natural way the near-symmetry around $x=1$ observed for the charge structure factor $S(q)$ (Fig.~\ref{fig:sq}). Since the magnetization depends on the number of spins, we do not expect it to have the same symmetry. However, the evaluation of the characteristic fields on the dumbbell model depends only on the possible \textit{charge} configurations (which should have $x=1$ as a symmetry point) and on the Zeeman energy of a spin.  This puts into question whether the curves in Fig.~\ref{fig:M2} may exhibit similar characteristic fields for values of $x$ equidistant to $x=1$.

The answer is at the same time trivial and deep. Any irreducible charge configuration at one side of $x=1$ does have a counterpart at the symmetrically related concentration. However, what makes a difference here are not the irreducible charges, but the vacuum where these charges exist. For small $x$, the neutral background is one able to produce more charges (and even, an ordered crystal made of them) with the magnetic field as a sort of chemical potential. For increasing $x$, the vacuum of irreducible charges becomes increasingly similar to a vacuum of spins, where no additional monopoles can be produced. In other words, although the structure factor for charges is symmetrical around $x=1$, nothing of the sort should be expected for the Fourier transform of the spin-spin correlation function.

\section{Summary and Conclusions}

In this work we have studied the behavior of spin-ice systems under doping with non-magnetic ions, concentrating on the simple dipolar model. Long range dipolar effects are not easily unveiled in regular experiments on spin ices; here we focus on certain features which are made evident only by the presence of magnetic dilution. In order to facilitate its eventual detection in real experiments, we used in our simulations single spin-flip dynamics and measurement protocols which parallel laboratory conditions. 

Due to dilution, fractional magnetic charges will be present in the ground state of the system; their abundance peaks at $x=1$ at the relatively high value of one magnetic charge every two tetrahedra. In spite of this, no long range charge order is found. Instead, the structure factor shows diffuse scattering indicative of local order; this may partially explain the reduction in the residual entropy found in previous experimental and theoretical works.  We examined the magnetization behavior under an applied magnetic field in the [111] crystallographic direction.  Our simulations show that the magnetization curves as a function of the magnetic field have markedly different characteristics than previous numerical studies, which are limited to the nearest neighbor model. The inclusion of dipolar interactions smooths the curves, rather than introducing additional jumps and plateaux, and better aligns with existing experimental data.  We have analyzed this problem using the dumbbell model, where magnetic moments are represented as pairs of oppositely charged magnetic monopoles.  We began by calculating the critical and spinodal fields corresponding to the phase transition observed in pure samples, and later introduced the method of virtual charges to account for magnetic defects.  In addition to providing a method to calculate characteristic fields, which we did for the case of isolated defects, it gives a qualitative picture that explains the smooth behavior of the simulated magnetization curves in terms of the variety of possible local environments created by the presence of random defects and long-range interactions.

\begin{acknowledgements}
R. A. B. would like to acknowledge useful discussions with M. Gingras during a stay invited by the ENS de Lyon. 
\end{acknowledgements}

\bibliography{ref}

\begin{thebibliography}{36}%
\makeatletter
\providecommand \@ifxundefined [1]{%
 \@ifx{#1\undefined}
}%
\providecommand \@ifnum [1]{%
 \ifnum #1\expandafter \@firstoftwo
 \else \expandafter \@secondoftwo
 \fi
}%
\providecommand \@ifx [1]{%
 \ifx #1\expandafter \@firstoftwo
 \else \expandafter \@secondoftwo
 \fi
}%
\providecommand \natexlab [1]{#1}%
\providecommand \enquote  [1]{``#1''}%
\providecommand \bibnamefont  [1]{#1}%
\providecommand \bibfnamefont [1]{#1}%
\providecommand \citenamefont [1]{#1}%
\providecommand \href@noop [0]{\@secondoftwo}%
\providecommand \href [0]{\begingroup \@sanitize@url \@href}%
\providecommand \@href[1]{\@@startlink{#1}\@@href}%
\providecommand \@@href[1]{\endgroup#1\@@endlink}%
\providecommand \@sanitize@url [0]{\catcode `\\12\catcode `\$12\catcode `\&12\catcode `\#12\catcode `\^12\catcode `\_12\catcode `\%12\relax}%
\providecommand \@@startlink[1]{}%
\providecommand \@@endlink[0]{}%
\providecommand \url  [0]{\begingroup\@sanitize@url \@url }%
\providecommand \@url [1]{\endgroup\@href {#1}{\urlprefix }}%
\providecommand \urlprefix  [0]{URL }%
\providecommand \Eprint [0]{\href }%
\providecommand \doibase [0]{https://doi.org/}%
\providecommand \selectlanguage [0]{\@gobble}%
\providecommand \bibinfo  [0]{\@secondoftwo}%
\providecommand \bibfield  [0]{\@secondoftwo}%
\providecommand \translation [1]{[#1]}%
\providecommand \BibitemOpen [0]{}%
\providecommand \bibitemStop [0]{}%
\providecommand \bibitemNoStop [0]{.\EOS\space}%
\providecommand \EOS [0]{\spacefactor3000\relax}%
\providecommand \BibitemShut  [1]{\csname bibitem#1\endcsname}%
\let\auto@bib@innerbib\@empty
\bibitem [{\citenamefont {Bramwell}\ and\ \citenamefont {Harris}(2020)}]{bramwell_history_spin_ice_2020}%
  \BibitemOpen
  \bibfield  {author} {\bibinfo {author} {\bibfnamefont {S.~T.}\ \bibnamefont {Bramwell}}\ and\ \bibinfo {author} {\bibfnamefont {M.~J.}\ \bibnamefont {Harris}},\ }\bibfield  {title} {\bibinfo {title} {The history of spin ice},\ }\href {https://doi.org/10.1088/1361-648X/ab8423} {\bibfield  {journal} {\bibinfo  {journal} {Journal of Physics: Condensed Matter}\ }\textbf {\bibinfo {volume} {32}},\ \bibinfo {pages} {374010} (\bibinfo {year} {2020})}\BibitemShut {NoStop}%
\bibitem [{\citenamefont {Snyder}\ \emph {et~al.}(2001)\citenamefont {Snyder}, \citenamefont {Slusky}, \citenamefont {Cava},\ and\ \citenamefont {Schiffer}}]{snyder2001spin}%
  \BibitemOpen
  \bibfield  {author} {\bibinfo {author} {\bibfnamefont {J.}~\bibnamefont {Snyder}}, \bibinfo {author} {\bibfnamefont {J.}~\bibnamefont {Slusky}}, \bibinfo {author} {\bibfnamefont {R.}~\bibnamefont {Cava}},\ and\ \bibinfo {author} {\bibfnamefont {P.}~\bibnamefont {Schiffer}},\ }\bibfield  {title} {\bibinfo {title} {How ‘spin ice’freezes},\ }\href@noop {} {\bibfield  {journal} {\bibinfo  {journal} {Nature}\ }\textbf {\bibinfo {volume} {413}},\ \bibinfo {pages} {48} (\bibinfo {year} {2001})}\BibitemShut {NoStop}%
\bibitem [{\citenamefont {Samarakoon}\ \emph {et~al.}(2022)\citenamefont {Samarakoon}, \citenamefont {Sokolowski}, \citenamefont {Klemke}, \citenamefont {Feyerherm}, \citenamefont {Meissner}, \citenamefont {Borzi}, \citenamefont {Ye}, \citenamefont {Zhang}, \citenamefont {Dun}, \citenamefont {Zhou} \emph {et~al.}}]{samarakoon2022structural}%
  \BibitemOpen
  \bibfield  {author} {\bibinfo {author} {\bibfnamefont {A.~M.}\ \bibnamefont {Samarakoon}}, \bibinfo {author} {\bibfnamefont {A.}~\bibnamefont {Sokolowski}}, \bibinfo {author} {\bibfnamefont {B.}~\bibnamefont {Klemke}}, \bibinfo {author} {\bibfnamefont {R.}~\bibnamefont {Feyerherm}}, \bibinfo {author} {\bibfnamefont {M.}~\bibnamefont {Meissner}}, \bibinfo {author} {\bibfnamefont {R.~A.}\ \bibnamefont {Borzi}}, \bibinfo {author} {\bibfnamefont {F.}~\bibnamefont {Ye}}, \bibinfo {author} {\bibfnamefont {Q.}~\bibnamefont {Zhang}}, \bibinfo {author} {\bibfnamefont {Z.}~\bibnamefont {Dun}}, \bibinfo {author} {\bibfnamefont {H.}~\bibnamefont {Zhou}}, \emph {et~al.},\ }\bibfield  {title} {\bibinfo {title} {Structural magnetic glassiness in the spin ice dy 2 ti 2 o 7},\ }\href@noop {} {\bibfield  {journal} {\bibinfo  {journal} {Physical Review Research}\ }\textbf {\bibinfo {volume} {4}},\ \bibinfo {pages} {033159} (\bibinfo {year} {2022})}\BibitemShut {NoStop}%
\bibitem [{\citenamefont {Hall{\'e}n}\ \emph {et~al.}(2022)\citenamefont {Hall{\'e}n}, \citenamefont {Grigera}, \citenamefont {Tennant}, \citenamefont {Castelnovo},\ and\ \citenamefont {Moessner}}]{hallen2022dynamical}%
  \BibitemOpen
  \bibfield  {author} {\bibinfo {author} {\bibfnamefont {J.~N.}\ \bibnamefont {Hall{\'e}n}}, \bibinfo {author} {\bibfnamefont {S.~A.}\ \bibnamefont {Grigera}}, \bibinfo {author} {\bibfnamefont {D.~A.}\ \bibnamefont {Tennant}}, \bibinfo {author} {\bibfnamefont {C.}~\bibnamefont {Castelnovo}},\ and\ \bibinfo {author} {\bibfnamefont {R.}~\bibnamefont {Moessner}},\ }\bibfield  {title} {\bibinfo {title} {Dynamical fractal and anomalous noise in a clean magnetic crystal},\ }\href@noop {} {\bibfield  {journal} {\bibinfo  {journal} {Science}\ }\textbf {\bibinfo {volume} {378}},\ \bibinfo {pages} {1218} (\bibinfo {year} {2022})}\BibitemShut {NoStop}%
\bibitem [{\citenamefont {Sen}\ and\ \citenamefont {Moessner}(2015)}]{sen2015topological}%
  \BibitemOpen
  \bibfield  {author} {\bibinfo {author} {\bibfnamefont {A.}~\bibnamefont {Sen}}\ and\ \bibinfo {author} {\bibfnamefont {R.}~\bibnamefont {Moessner}},\ }\bibfield  {title} {\bibinfo {title} {Topological spin glass in diluted spin ice},\ }\href@noop {} {\bibfield  {journal} {\bibinfo  {journal} {Physical review letters}\ }\textbf {\bibinfo {volume} {114}},\ \bibinfo {pages} {247207} (\bibinfo {year} {2015})}\BibitemShut {NoStop}%
\bibitem [{\citenamefont {Scharffe}\ \emph {et~al.}(2015)\citenamefont {Scharffe}, \citenamefont {Breunig}, \citenamefont {Cho}, \citenamefont {Laschitzky}, \citenamefont {Valldor}, \citenamefont {Welter},\ and\ \citenamefont {Lorenz}}]{scharffe2015suppression}%
  \BibitemOpen
  \bibfield  {author} {\bibinfo {author} {\bibfnamefont {S.}~\bibnamefont {Scharffe}}, \bibinfo {author} {\bibfnamefont {O.}~\bibnamefont {Breunig}}, \bibinfo {author} {\bibfnamefont {V.}~\bibnamefont {Cho}}, \bibinfo {author} {\bibfnamefont {P.}~\bibnamefont {Laschitzky}}, \bibinfo {author} {\bibfnamefont {M.}~\bibnamefont {Valldor}}, \bibinfo {author} {\bibfnamefont {J.}~\bibnamefont {Welter}},\ and\ \bibinfo {author} {\bibfnamefont {T.}~\bibnamefont {Lorenz}},\ }\bibfield  {title} {\bibinfo {title} {Suppression of pauling's residual entropy in the dilute spin ice (dy 1- x y x) 2 ti 2 o 7},\ }\href@noop {} {\bibfield  {journal} {\bibinfo  {journal} {Physical Review B}\ }\textbf {\bibinfo {volume} {92}},\ \bibinfo {pages} {180405} (\bibinfo {year} {2015})}\BibitemShut {NoStop}%
\bibitem [{\citenamefont {Castelnovo}\ \emph {et~al.}(2008)\citenamefont {Castelnovo}, \citenamefont {Moessner},\ and\ \citenamefont {Sondhi}}]{castelnovo_2008}%
  \BibitemOpen
  \bibfield  {author} {\bibinfo {author} {\bibfnamefont {C.}~\bibnamefont {Castelnovo}}, \bibinfo {author} {\bibfnamefont {R.}~\bibnamefont {Moessner}},\ and\ \bibinfo {author} {\bibfnamefont {S.~L.}\ \bibnamefont {Sondhi}},\ }\bibfield  {title} {\bibinfo {title} {Magnetic monopoles in spin ice},\ }\href {https://doi.org/10.1038/nature06433} {\bibfield  {journal} {\bibinfo  {journal} {Nature}\ }\textbf {\bibinfo {volume} {451}},\ \bibinfo {pages} {42} (\bibinfo {year} {2008})}\BibitemShut {NoStop}%
\bibitem [{\citenamefont {Snyder}\ \emph {et~al.}(2004{\natexlab{a}})\citenamefont {Snyder}, \citenamefont {Ueland}, \citenamefont {Mizel}, \citenamefont {Slusky}, \citenamefont {Karunadasa}, \citenamefont {Cava},\ and\ \citenamefont {Schiffer}}]{snyder2004quantum}%
  \BibitemOpen
  \bibfield  {author} {\bibinfo {author} {\bibfnamefont {J.}~\bibnamefont {Snyder}}, \bibinfo {author} {\bibfnamefont {B.}~\bibnamefont {Ueland}}, \bibinfo {author} {\bibfnamefont {A.}~\bibnamefont {Mizel}}, \bibinfo {author} {\bibfnamefont {J.}~\bibnamefont {Slusky}}, \bibinfo {author} {\bibfnamefont {H.}~\bibnamefont {Karunadasa}}, \bibinfo {author} {\bibfnamefont {R.}~\bibnamefont {Cava}},\ and\ \bibinfo {author} {\bibfnamefont {P.}~\bibnamefont {Schiffer}},\ }\bibfield  {title} {\bibinfo {title} {Quantum and thermal spin relaxation in the diluted spin ice dy 2- x m x ti 2 o 7 (m= lu, y)},\ }\href@noop {} {\bibfield  {journal} {\bibinfo  {journal} {Physical Review B—Condensed Matter and Materials Physics}\ }\textbf {\bibinfo {volume} {70}},\ \bibinfo {pages} {184431} (\bibinfo {year} {2004}{\natexlab{a}})}\BibitemShut {NoStop}%
\bibitem [{\citenamefont {Lin}\ \emph {et~al.}(2014)\citenamefont {Lin}, \citenamefont {Ke}, \citenamefont {Thesberg}, \citenamefont {Schiffer}, \citenamefont {Melko},\ and\ \citenamefont {Gingras}}]{lin2014nonmonotonic}%
  \BibitemOpen
  \bibfield  {author} {\bibinfo {author} {\bibfnamefont {T.}~\bibnamefont {Lin}}, \bibinfo {author} {\bibfnamefont {X.}~\bibnamefont {Ke}}, \bibinfo {author} {\bibfnamefont {M.}~\bibnamefont {Thesberg}}, \bibinfo {author} {\bibfnamefont {P.}~\bibnamefont {Schiffer}}, \bibinfo {author} {\bibfnamefont {R.}~\bibnamefont {Melko}},\ and\ \bibinfo {author} {\bibfnamefont {M.}~\bibnamefont {Gingras}},\ }\bibfield  {title} {\bibinfo {title} {Nonmonotonic residual entropy in diluted spin ice: A comparison between monte carlo simulations of diluted dipolar spin ice models and experimental results},\ }\href@noop {} {\bibfield  {journal} {\bibinfo  {journal} {Physical Review B}\ }\textbf {\bibinfo {volume} {90}},\ \bibinfo {pages} {214433} (\bibinfo {year} {2014})}\BibitemShut {NoStop}%
\bibitem [{\citenamefont {Andresen}\ \emph {et~al.}(2014)\citenamefont {Andresen}, \citenamefont {Katzgraber}, \citenamefont {Oganesyan},\ and\ \citenamefont {Schechter}}]{AndPRX_2014}%
  \BibitemOpen
  \bibfield  {author} {\bibinfo {author} {\bibfnamefont {J.~C.}\ \bibnamefont {Andresen}}, \bibinfo {author} {\bibfnamefont {H.~G.}\ \bibnamefont {Katzgraber}}, \bibinfo {author} {\bibfnamefont {V.}~\bibnamefont {Oganesyan}},\ and\ \bibinfo {author} {\bibfnamefont {M.}~\bibnamefont {Schechter}},\ }\bibfield  {title} {\bibinfo {title} {Existence of a thermodynamic spin-glass phase in the zero-concentration limit of anisotropic dipolar systems},\ }\href {https://doi.org/10.1103/PhysRevX.4.041016} {\bibfield  {journal} {\bibinfo  {journal} {Phys. Rev. X}\ }\textbf {\bibinfo {volume} {4}},\ \bibinfo {pages} {041016} (\bibinfo {year} {2014})}\BibitemShut {NoStop}%
\bibitem [{\citenamefont {Prabhakaran}\ and\ \citenamefont {Boothroyd}(2011)}]{prabhakaran2011crystal}%
  \BibitemOpen
  \bibfield  {author} {\bibinfo {author} {\bibfnamefont {D.}~\bibnamefont {Prabhakaran}}\ and\ \bibinfo {author} {\bibfnamefont {A.}~\bibnamefont {Boothroyd}},\ }\bibfield  {title} {\bibinfo {title} {Crystal growth of spin-ice pyrochlores by the floating-zone method},\ }\href@noop {} {\bibfield  {journal} {\bibinfo  {journal} {Journal of Crystal Growth}\ }\textbf {\bibinfo {volume} {318}},\ \bibinfo {pages} {1053} (\bibinfo {year} {2011})}\BibitemShut {NoStop}%
\bibitem [{\citenamefont {Liu}\ \emph {et~al.}(2015)\citenamefont {Liu}, \citenamefont {Zou}, \citenamefont {Zhang}, \citenamefont {Zhang}, \citenamefont {Zhang},\ and\ \citenamefont {Zhang}}]{liu2015dy}%
  \BibitemOpen
  \bibfield  {author} {\bibinfo {author} {\bibfnamefont {H.}~\bibnamefont {Liu}}, \bibinfo {author} {\bibfnamefont {Y.-M.}\ \bibnamefont {Zou}}, \bibinfo {author} {\bibfnamefont {S.-L.}\ \bibnamefont {Zhang}}, \bibinfo {author} {\bibfnamefont {R.-R.}\ \bibnamefont {Zhang}}, \bibinfo {author} {\bibfnamefont {C.-J.}\ \bibnamefont {Zhang}},\ and\ \bibinfo {author} {\bibfnamefont {Y.-H.}\ \bibnamefont {Zhang}},\ }\bibfield  {title} {\bibinfo {title} {Dy 2- x y x ti 2 o 7: phonon vibration and magnetization with dilution},\ }\href@noop {} {\bibfield  {journal} {\bibinfo  {journal} {Rare Metals}\ }\textbf {\bibinfo {volume} {34}},\ \bibinfo {pages} {81} (\bibinfo {year} {2015})}\BibitemShut {NoStop}%
\bibitem [{\citenamefont {Peretyatko}\ \emph {et~al.}(2017)\citenamefont {Peretyatko}, \citenamefont {Nefedev},\ and\ \citenamefont {Okabe}}]{peretyatko2017interplay}%
  \BibitemOpen
  \bibfield  {author} {\bibinfo {author} {\bibfnamefont {A.}~\bibnamefont {Peretyatko}}, \bibinfo {author} {\bibfnamefont {K.}~\bibnamefont {Nefedev}},\ and\ \bibinfo {author} {\bibfnamefont {Y.}~\bibnamefont {Okabe}},\ }\bibfield  {title} {\bibinfo {title} {Interplay of dilution and magnetic field in the nearest-neighbor spin-ice model on the pyrochlore lattice},\ }\href@noop {} {\bibfield  {journal} {\bibinfo  {journal} {Physical Review B}\ }\textbf {\bibinfo {volume} {95}},\ \bibinfo {pages} {144410} (\bibinfo {year} {2017})}\BibitemShut {NoStop}%
\bibitem [{\citenamefont {Sakakibara}\ \emph {et~al.}(2003)\citenamefont {Sakakibara}, \citenamefont {Tayama}, \citenamefont {Hiroi}, \citenamefont {Matsuhira},\ and\ \citenamefont {Takagi}}]{sakakibara2003observation}%
  \BibitemOpen
  \bibfield  {author} {\bibinfo {author} {\bibfnamefont {T.}~\bibnamefont {Sakakibara}}, \bibinfo {author} {\bibfnamefont {T.}~\bibnamefont {Tayama}}, \bibinfo {author} {\bibfnamefont {Z.}~\bibnamefont {Hiroi}}, \bibinfo {author} {\bibfnamefont {K.}~\bibnamefont {Matsuhira}},\ and\ \bibinfo {author} {\bibfnamefont {S.}~\bibnamefont {Takagi}},\ }\bibfield  {title} {\bibinfo {title} {Observation of a liquid-gas-type transition in the pyrochlore spin ice compound d y 2 t i 2 o 7 in a magnetic field},\ }\href@noop {} {\bibfield  {journal} {\bibinfo  {journal} {Physical review letters}\ }\textbf {\bibinfo {volume} {90}},\ \bibinfo {pages} {207205} (\bibinfo {year} {2003})}\BibitemShut {NoStop}%
\bibitem [{\citenamefont {Isakov}\ \emph {et~al.}(2005)\citenamefont {Isakov}, \citenamefont {Moessner},\ and\ \citenamefont {Sondhi}}]{isakov_2005spin}%
  \BibitemOpen
  \bibfield  {author} {\bibinfo {author} {\bibfnamefont {S.~V.}\ \bibnamefont {Isakov}}, \bibinfo {author} {\bibfnamefont {R.}~\bibnamefont {Moessner}},\ and\ \bibinfo {author} {\bibfnamefont {S.~L.}\ \bibnamefont {Sondhi}},\ }\bibfield  {title} {\bibinfo {title} {Why spin ice obeys the ice rules},\ }\href@noop {} {\bibfield  {journal} {\bibinfo  {journal} {Physical review letters}\ }\textbf {\bibinfo {volume} {95}},\ \bibinfo {pages} {217201} (\bibinfo {year} {2005})}\BibitemShut {NoStop}%
\bibitem [{\citenamefont {Brooks-Bartlett}\ \emph {et~al.}(2014)\citenamefont {Brooks-Bartlett}, \citenamefont {Banks}, \citenamefont {Jaubert}, \citenamefont {Harman-Clarke},\ and\ \citenamefont {Holdsworth}}]{brooks_artificial_2014}%
  \BibitemOpen
  \bibfield  {author} {\bibinfo {author} {\bibfnamefont {M.~E.}\ \bibnamefont {Brooks-Bartlett}}, \bibinfo {author} {\bibfnamefont {S.~T.}\ \bibnamefont {Banks}}, \bibinfo {author} {\bibfnamefont {L.~D.~C.}\ \bibnamefont {Jaubert}}, \bibinfo {author} {\bibfnamefont {A.}~\bibnamefont {Harman-Clarke}},\ and\ \bibinfo {author} {\bibfnamefont {P.~C.~W.}\ \bibnamefont {Holdsworth}},\ }\bibfield  {title} {\bibinfo {title} {Magnetic-{Moment} {Fragmentation} and {Monopole} {Crystallization}},\ }\href {https://doi.org/10.1103/PhysRevX.4.011007} {\bibfield  {journal} {\bibinfo  {journal} {Physical Review X}\ }\textbf {\bibinfo {volume} {4}},\ \bibinfo {pages} {011007} (\bibinfo {year} {2014})}\BibitemShut {NoStop}%
\bibitem [{\citenamefont {Guruciaga}\ \emph {et~al.}(2014)\citenamefont {Guruciaga}, \citenamefont {Grigera},\ and\ \citenamefont {Borzi}}]{guruciaga2014monopole}%
  \BibitemOpen
  \bibfield  {author} {\bibinfo {author} {\bibfnamefont {P.~C.}\ \bibnamefont {Guruciaga}}, \bibinfo {author} {\bibfnamefont {S.~A.}\ \bibnamefont {Grigera}},\ and\ \bibinfo {author} {\bibfnamefont {R.~A.}\ \bibnamefont {Borzi}},\ }\bibfield  {title} {\bibinfo {title} {Monopole ordered phases in dipolar and nearest-neighbors ising pyrochlore: From spin ice to the all-in--all-out antiferromagnet},\ }\href@noop {} {\bibfield  {journal} {\bibinfo  {journal} {Physical Review B}\ }\textbf {\bibinfo {volume} {90}},\ \bibinfo {pages} {184423} (\bibinfo {year} {2014})}\BibitemShut {NoStop}%
\bibitem [{\citenamefont {Yavors’~kii}\ \emph {et~al.}(2008)\citenamefont {Yavors’~kii}, \citenamefont {Fennell}, \citenamefont {Gingras},\ and\ \citenamefont {Bramwell}}]{yavors2008dy}%
  \BibitemOpen
  \bibfield  {author} {\bibinfo {author} {\bibfnamefont {T.}~\bibnamefont {Yavors’~kii}}, \bibinfo {author} {\bibfnamefont {T.}~\bibnamefont {Fennell}}, \bibinfo {author} {\bibfnamefont {M.~J.}\ \bibnamefont {Gingras}},\ and\ \bibinfo {author} {\bibfnamefont {S.~T.}\ \bibnamefont {Bramwell}},\ }\bibfield  {title} {\bibinfo {title} {Dy 2 ti 2 o 7 spin ice: a test case for emergent clusters in a frustrated magnet},\ }\href@noop {} {\bibfield  {journal} {\bibinfo  {journal} {Physical review letters}\ }\textbf {\bibinfo {volume} {101}},\ \bibinfo {pages} {037204} (\bibinfo {year} {2008})}\BibitemShut {NoStop}%
\bibitem [{\citenamefont {Borzi}\ \emph {et~al.}(2016)\citenamefont {Borzi}, \citenamefont {Gómez~Albarracín}, \citenamefont {Rosales}, \citenamefont {Rossini}, \citenamefont {Steppke}, \citenamefont {Prabhakaran}, \citenamefont {Mackenzie}, \citenamefont {Cabra},\ and\ \citenamefont {Grigera}}]{Borzi2016}%
  \BibitemOpen
  \bibfield  {author} {\bibinfo {author} {\bibfnamefont {R.~A.}\ \bibnamefont {Borzi}}, \bibinfo {author} {\bibfnamefont {F.~A.}\ \bibnamefont {Gómez~Albarracín}}, \bibinfo {author} {\bibfnamefont {H.~D.}\ \bibnamefont {Rosales}}, \bibinfo {author} {\bibfnamefont {G.~L.}\ \bibnamefont {Rossini}}, \bibinfo {author} {\bibfnamefont {A.}~\bibnamefont {Steppke}}, \bibinfo {author} {\bibfnamefont {D.}~\bibnamefont {Prabhakaran}}, \bibinfo {author} {\bibfnamefont {A.~P.}\ \bibnamefont {Mackenzie}}, \bibinfo {author} {\bibfnamefont {D.~C.}\ \bibnamefont {Cabra}},\ and\ \bibinfo {author} {\bibfnamefont {S.~A.}\ \bibnamefont {Grigera}},\ }\bibfield  {title} {\bibinfo {title} {Intermediate magnetization state and competing orders in dy2ti2o7 and ho2ti2o7},\ }\href {https://doi.org/10.1038/ncomms12592} {\bibfield  {journal} {\bibinfo  {journal} {Nature Communications}\ }\textbf {\bibinfo {volume} {7}},\ \bibinfo {pages} {12592} (\bibinfo {year} {2016})}\BibitemShut {NoStop}%
\bibitem [{\citenamefont {Henelius}\ \emph {et~al.}(2016)\citenamefont {Henelius}, \citenamefont {Lin}, \citenamefont {Enjalran}, \citenamefont {Hao}, \citenamefont {Rau}, \citenamefont {Altosaar}, \citenamefont {Flicker}, \citenamefont {Yavors'~Kii},\ and\ \citenamefont {Gingras}}]{henelius2016refrustration}%
  \BibitemOpen
  \bibfield  {author} {\bibinfo {author} {\bibfnamefont {P.}~\bibnamefont {Henelius}}, \bibinfo {author} {\bibfnamefont {T.}~\bibnamefont {Lin}}, \bibinfo {author} {\bibfnamefont {M.}~\bibnamefont {Enjalran}}, \bibinfo {author} {\bibfnamefont {Z.}~\bibnamefont {Hao}}, \bibinfo {author} {\bibfnamefont {J.}~\bibnamefont {Rau}}, \bibinfo {author} {\bibfnamefont {J.}~\bibnamefont {Altosaar}}, \bibinfo {author} {\bibfnamefont {F.}~\bibnamefont {Flicker}}, \bibinfo {author} {\bibfnamefont {T.}~\bibnamefont {Yavors'~Kii}},\ and\ \bibinfo {author} {\bibfnamefont {M.}~\bibnamefont {Gingras}},\ }\bibfield  {title} {\bibinfo {title} {Refrustration and competing orders in the prototypical dy 2 ti 2 o 7 spin ice material},\ }\href@noop {} {\bibfield  {journal} {\bibinfo  {journal} {Physical Review B}\ }\textbf {\bibinfo {volume} {93}},\ \bibinfo {pages} {024402} (\bibinfo {year} {2016})}\BibitemShut {NoStop}%
\bibitem [{\citenamefont {Samarakoon}\ \emph {et~al.}(2020)\citenamefont {Samarakoon}, \citenamefont {Barros}, \citenamefont {Li}, \citenamefont {Eisenbach}, \citenamefont {Zhang}, \citenamefont {Ye}, \citenamefont {Sharma}, \citenamefont {Dun}, \citenamefont {Zhou}, \citenamefont {Grigera} \emph {et~al.}}]{samarakoon2020machine}%
  \BibitemOpen
  \bibfield  {author} {\bibinfo {author} {\bibfnamefont {A.~M.}\ \bibnamefont {Samarakoon}}, \bibinfo {author} {\bibfnamefont {K.}~\bibnamefont {Barros}}, \bibinfo {author} {\bibfnamefont {Y.~W.}\ \bibnamefont {Li}}, \bibinfo {author} {\bibfnamefont {M.}~\bibnamefont {Eisenbach}}, \bibinfo {author} {\bibfnamefont {Q.}~\bibnamefont {Zhang}}, \bibinfo {author} {\bibfnamefont {F.}~\bibnamefont {Ye}}, \bibinfo {author} {\bibfnamefont {V.}~\bibnamefont {Sharma}}, \bibinfo {author} {\bibfnamefont {Z.}~\bibnamefont {Dun}}, \bibinfo {author} {\bibfnamefont {H.}~\bibnamefont {Zhou}}, \bibinfo {author} {\bibfnamefont {S.~A.}\ \bibnamefont {Grigera}}, \emph {et~al.},\ }\bibfield  {title} {\bibinfo {title} {Machine-learning-assisted insight into spin ice dy2ti2o7},\ }\href@noop {} {\bibfield  {journal} {\bibinfo  {journal} {Nature communications}\ }\textbf {\bibinfo {volume} {11}},\ \bibinfo {pages} {892} (\bibinfo {year} {2020})}\BibitemShut {NoStop}%
\bibitem [{\citenamefont {Melko}\ and\ \citenamefont {Gingras}(2004)}]{melko2004monte}%
  \BibitemOpen
  \bibfield  {author} {\bibinfo {author} {\bibfnamefont {R.~G.}\ \bibnamefont {Melko}}\ and\ \bibinfo {author} {\bibfnamefont {M.~J.}\ \bibnamefont {Gingras}},\ }\bibfield  {title} {\bibinfo {title} {Monte carlo studies of the dipolar spin ice model},\ }\href@noop {} {\bibfield  {journal} {\bibinfo  {journal} {Journal of Physics: Condensed Matter}\ }\textbf {\bibinfo {volume} {16}},\ \bibinfo {pages} {R1277} (\bibinfo {year} {2004})}\BibitemShut {NoStop}%
\bibitem [{\citenamefont {Borzi}\ \emph {et~al.}(2013)\citenamefont {Borzi}, \citenamefont {Slobinsky},\ and\ \citenamefont {Grigera}}]{borzi_2013}%
  \BibitemOpen
  \bibfield  {author} {\bibinfo {author} {\bibfnamefont {R.~A.}\ \bibnamefont {Borzi}}, \bibinfo {author} {\bibfnamefont {D.}~\bibnamefont {Slobinsky}},\ and\ \bibinfo {author} {\bibfnamefont {S.~A.}\ \bibnamefont {Grigera}},\ }\bibfield  {title} {\bibinfo {title} {Charge {{Ordering}} in a {{Pure Spin Model}}: {{Dipolar Spin Ice}}},\ }\href {https://doi.org/10.1103/PhysRevLett.111.147204} {\bibfield  {journal} {\bibinfo  {journal} {Physical Review Letters}\ }\textbf {\bibinfo {volume} {111}},\ \bibinfo {pages} {147204} (\bibinfo {year} {2013})}\BibitemShut {NoStop}%
\bibitem [{\citenamefont {Matsuhira}\ \emph {et~al.}(2000)\citenamefont {Matsuhira}, \citenamefont {Hinatsu}, \citenamefont {Tenya},\ and\ \citenamefont {Sakakibara}}]{matsuhira2000low}%
  \BibitemOpen
  \bibfield  {author} {\bibinfo {author} {\bibfnamefont {K.}~\bibnamefont {Matsuhira}}, \bibinfo {author} {\bibfnamefont {Y.}~\bibnamefont {Hinatsu}}, \bibinfo {author} {\bibfnamefont {K.}~\bibnamefont {Tenya}},\ and\ \bibinfo {author} {\bibfnamefont {T.}~\bibnamefont {Sakakibara}},\ }\bibfield  {title} {\bibinfo {title} {Low temperature magnetic properties of frustrated pyrochlore ferromagnets ho2sn2o7 and ho2ti2o7},\ }\href@noop {} {\bibfield  {journal} {\bibinfo  {journal} {Journal of Physics: Condensed Matter}\ }\textbf {\bibinfo {volume} {12}},\ \bibinfo {pages} {L649} (\bibinfo {year} {2000})}\BibitemShut {NoStop}%
\bibitem [{\citenamefont {Snyder}\ \emph {et~al.}(2004{\natexlab{b}})\citenamefont {Snyder}, \citenamefont {Ueland}, \citenamefont {Slusky}, \citenamefont {Karunadasa}, \citenamefont {Cava},\ and\ \citenamefont {Schiffer}}]{snyder2004low}%
  \BibitemOpen
  \bibfield  {author} {\bibinfo {author} {\bibfnamefont {J.}~\bibnamefont {Snyder}}, \bibinfo {author} {\bibfnamefont {B.}~\bibnamefont {Ueland}}, \bibinfo {author} {\bibfnamefont {J.}~\bibnamefont {Slusky}}, \bibinfo {author} {\bibfnamefont {H.}~\bibnamefont {Karunadasa}}, \bibinfo {author} {\bibfnamefont {R.}~\bibnamefont {Cava}},\ and\ \bibinfo {author} {\bibfnamefont {P.}~\bibnamefont {Schiffer}},\ }\bibfield  {title} {\bibinfo {title} {Low-temperature spin freezing in the dy 2 ti 2 o 7 spin ice},\ }\href@noop {} {\bibfield  {journal} {\bibinfo  {journal} {Physical Review B}\ }\textbf {\bibinfo {volume} {69}},\ \bibinfo {pages} {064414} (\bibinfo {year} {2004}{\natexlab{b}})}\BibitemShut {NoStop}%
\bibitem [{\citenamefont {Slobinsky}\ \emph {et~al.}(2019)\citenamefont {Slobinsky}, \citenamefont {Pili},\ and\ \citenamefont {Borzi}}]{slobinsky_field_2019}%
  \BibitemOpen
  \bibfield  {author} {\bibinfo {author} {\bibfnamefont {D.}~\bibnamefont {Slobinsky}}, \bibinfo {author} {\bibfnamefont {L.}~\bibnamefont {Pili}},\ and\ \bibinfo {author} {\bibfnamefont {R.~A.}\ \bibnamefont {Borzi}},\ }\bibfield  {title} {\bibinfo {title} {Polarized monopole liquid: {A} {Coulomb} phase in a fluid of magnetic charges},\ }\href {https://doi.org/10.1103/PhysRevB.100.020405} {\bibfield  {journal} {\bibinfo  {journal} {Physical Review B}\ }\textbf {\bibinfo {volume} {100}},\ \bibinfo {pages} {020405} (\bibinfo {year} {2019})}\BibitemShut {NoStop}%
\bibitem [{\citenamefont {Slobinsky}\ \emph {et~al.}(2021)\citenamefont {Slobinsky}, \citenamefont {Pili}, \citenamefont {Baglietto}, \citenamefont {Grigera},\ and\ \citenamefont {Borzi}}]{slobinsky_monopole_2021}%
  \BibitemOpen
  \bibfield  {author} {\bibinfo {author} {\bibfnamefont {D.}~\bibnamefont {Slobinsky}}, \bibinfo {author} {\bibfnamefont {L.}~\bibnamefont {Pili}}, \bibinfo {author} {\bibfnamefont {G.}~\bibnamefont {Baglietto}}, \bibinfo {author} {\bibfnamefont {S.~A.}\ \bibnamefont {Grigera}},\ and\ \bibinfo {author} {\bibfnamefont {R.~A.}\ \bibnamefont {Borzi}},\ }\bibfield  {title} {\bibinfo {title} {Monopole matter from magnetoelastic coupling in the {Ising} pyrochlore},\ }\href {https://doi.org/10.1038/s42005-021-00552-0} {\bibfield  {journal} {\bibinfo  {journal} {Communications Physics}\ }\textbf {\bibinfo {volume} {4}},\ \bibinfo {pages} {56} (\bibinfo {year} {2021})}\BibitemShut {NoStop}%
\bibitem [{\citenamefont {Slobinsky}\ \emph {et~al.}(2018)\citenamefont {Slobinsky}, \citenamefont {Baglietto},\ and\ \citenamefont {Borzi}}]{slobinsky2018}%
  \BibitemOpen
  \bibfield  {author} {\bibinfo {author} {\bibfnamefont {D.}~\bibnamefont {Slobinsky}}, \bibinfo {author} {\bibfnamefont {G.}~\bibnamefont {Baglietto}},\ and\ \bibinfo {author} {\bibfnamefont {R.~A.}\ \bibnamefont {Borzi}},\ }\bibfield  {title} {\bibinfo {title} {Charge and spin correlations in the monopole liquid},\ }\href {https://doi.org/10.1103/PhysRevB.97.174422} {\bibfield  {journal} {\bibinfo  {journal} {Physical Review B}\ }\textbf {\bibinfo {volume} {97}},\ \bibinfo {pages} {174422} (\bibinfo {year} {2018})}\BibitemShut {NoStop}%
\bibitem [{\citenamefont {Siddharthan}\ \emph {et~al.}(2001)\citenamefont {Siddharthan}, \citenamefont {Shastry},\ and\ \citenamefont {Ramirez}}]{siddharthan2001spin}%
  \BibitemOpen
  \bibfield  {author} {\bibinfo {author} {\bibfnamefont {R.}~\bibnamefont {Siddharthan}}, \bibinfo {author} {\bibfnamefont {B.}~\bibnamefont {Shastry}},\ and\ \bibinfo {author} {\bibfnamefont {A.}~\bibnamefont {Ramirez}},\ }\bibfield  {title} {\bibinfo {title} {Spin ordering and partial ordering in holmium titanate and related systems},\ }\href@noop {} {\bibfield  {journal} {\bibinfo  {journal} {Physical Review B}\ }\textbf {\bibinfo {volume} {63}},\ \bibinfo {pages} {184412} (\bibinfo {year} {2001})}\BibitemShut {NoStop}%
\bibitem [{\citenamefont {den Hertog}\ and\ \citenamefont {Gingras}(2000)}]{den2000dipolar}%
  \BibitemOpen
  \bibfield  {author} {\bibinfo {author} {\bibfnamefont {B.~C.}\ \bibnamefont {den Hertog}}\ and\ \bibinfo {author} {\bibfnamefont {M.~J.}\ \bibnamefont {Gingras}},\ }\bibfield  {title} {\bibinfo {title} {Dipolar interactions and origin of spin ice in ising pyrochlore magnets},\ }\href@noop {} {\bibfield  {journal} {\bibinfo  {journal} {Physical review letters}\ }\textbf {\bibinfo {volume} {84}},\ \bibinfo {pages} {3430} (\bibinfo {year} {2000})}\BibitemShut {NoStop}%
\bibitem [{\citenamefont {Gaunt}\ and\ \citenamefont {Sykes}(1983)}]{gaunt1983series}%
  \BibitemOpen
  \bibfield  {author} {\bibinfo {author} {\bibfnamefont {D.}~\bibnamefont {Gaunt}}\ and\ \bibinfo {author} {\bibfnamefont {M.}~\bibnamefont {Sykes}},\ }\bibfield  {title} {\bibinfo {title} {Series study of random percolation in three dimensions},\ }\href@noop {} {\bibfield  {journal} {\bibinfo  {journal} {Journal of Physics A: Mathematical and General}\ }\textbf {\bibinfo {volume} {16}},\ \bibinfo {pages} {783} (\bibinfo {year} {1983})}\BibitemShut {NoStop}%
\bibitem [{\citenamefont {Ke}\ \emph {et~al.}(2007)\citenamefont {Ke}, \citenamefont {Freitas}, \citenamefont {Ueland}, \citenamefont {Lau}, \citenamefont {Dahlberg}, \citenamefont {Cava}, \citenamefont {Moessner},\ and\ \citenamefont {Schiffer}}]{ke2007nonmonotonic}%
  \BibitemOpen
  \bibfield  {author} {\bibinfo {author} {\bibfnamefont {X.}~\bibnamefont {Ke}}, \bibinfo {author} {\bibfnamefont {R.}~\bibnamefont {Freitas}}, \bibinfo {author} {\bibfnamefont {B.}~\bibnamefont {Ueland}}, \bibinfo {author} {\bibfnamefont {G.}~\bibnamefont {Lau}}, \bibinfo {author} {\bibfnamefont {M.}~\bibnamefont {Dahlberg}}, \bibinfo {author} {\bibfnamefont {R.}~\bibnamefont {Cava}}, \bibinfo {author} {\bibfnamefont {R.}~\bibnamefont {Moessner}},\ and\ \bibinfo {author} {\bibfnamefont {P.}~\bibnamefont {Schiffer}},\ }\bibfield  {title} {\bibinfo {title} {Nonmonotonic zero-point entropy in diluted spin ice},\ }\href@noop {} {\bibfield  {journal} {\bibinfo  {journal} {Physical review letters}\ }\textbf {\bibinfo {volume} {99}},\ \bibinfo {pages} {137203} (\bibinfo {year} {2007})}\BibitemShut {NoStop}%
\bibitem [{\citenamefont {Aoki}\ \emph {et~al.}(2004)\citenamefont {Aoki}, \citenamefont {Sakakibara}, \citenamefont {Matsuhira},\ and\ \citenamefont {Hiroi}}]{aoki2004magnetocaloric}%
  \BibitemOpen
  \bibfield  {author} {\bibinfo {author} {\bibfnamefont {H.}~\bibnamefont {Aoki}}, \bibinfo {author} {\bibfnamefont {T.}~\bibnamefont {Sakakibara}}, \bibinfo {author} {\bibfnamefont {K.}~\bibnamefont {Matsuhira}},\ and\ \bibinfo {author} {\bibfnamefont {Z.}~\bibnamefont {Hiroi}},\ }\bibfield  {title} {\bibinfo {title} {Magnetocaloric effect study on the pyrochlore spin ice compound dy2ti2o7 in a [111] magnetic field},\ }\href@noop {} {\bibfield  {journal} {\bibinfo  {journal} {Journal of the Physical Society of Japan}\ }\textbf {\bibinfo {volume} {73}},\ \bibinfo {pages} {2851} (\bibinfo {year} {2004})}\BibitemShut {NoStop}%
\bibitem [{\citenamefont {Matsuhira}\ \emph {et~al.}(2002)\citenamefont {Matsuhira}, \citenamefont {Hiroi}, \citenamefont {Tayama}, \citenamefont {Takagi},\ and\ \citenamefont {Sakakibara}}]{matsuhira_field_2002}%
  \BibitemOpen
  \bibfield  {author} {\bibinfo {author} {\bibfnamefont {K.}~\bibnamefont {Matsuhira}}, \bibinfo {author} {\bibfnamefont {Z.}~\bibnamefont {Hiroi}}, \bibinfo {author} {\bibfnamefont {T.}~\bibnamefont {Tayama}}, \bibinfo {author} {\bibfnamefont {S.}~\bibnamefont {Takagi}},\ and\ \bibinfo {author} {\bibfnamefont {T.}~\bibnamefont {Sakakibara}},\ }\bibfield  {title} {\bibinfo {title} {A new macroscopically degenerate ground state in the spin ice compound {Dy} $_{\textrm{2}}$ {Ti} $_{\textrm{2}}$ {O} $_{\textrm{7}}$ under a magnetic field},\ }\href {https://doi.org/10.1088/0953-8984/14/29/101} {\bibfield  {journal} {\bibinfo  {journal} {Journal of Physics: Condensed Matter}\ }\textbf {\bibinfo {volume} {14}},\ \bibinfo {pages} {L559} (\bibinfo {year} {2002})}\BibitemShut {NoStop}%
\bibitem [{\citenamefont {Higashinaka}\ \emph {et~al.}(2003)\citenamefont {Higashinaka}, \citenamefont {Fukazawa},\ and\ \citenamefont {Maeno}}]{higashinaka2003anisotropic}%
  \BibitemOpen
  \bibfield  {author} {\bibinfo {author} {\bibfnamefont {R.}~\bibnamefont {Higashinaka}}, \bibinfo {author} {\bibfnamefont {H.}~\bibnamefont {Fukazawa}},\ and\ \bibinfo {author} {\bibfnamefont {Y.}~\bibnamefont {Maeno}},\ }\bibfield  {title} {\bibinfo {title} {Anisotropic release of the residual zero-point entropy in the spin ice compound dy 2 ti 2 o 7: Kagome ice behavior},\ }\href@noop {} {\bibfield  {journal} {\bibinfo  {journal} {Physical Review B}\ }\textbf {\bibinfo {volume} {68}},\ \bibinfo {pages} {014415} (\bibinfo {year} {2003})}\BibitemShut {NoStop}%
\bibitem [{Note1()}]{Note1}%
  \BibitemOpen
  \bibinfo {note} {One alternative way to see this is to consider a string of flipped spins connecting one charge in Fig.~\ref {fig:defects} bottom left to the other (the ``open loops'' of ref.~\cite {lin2014nonmonotonic}). Another one is to see it a flipping the \protect \textit {ghost spin} of Ref.~\cite {sen2015topological}. It is easy to see that on introducing such a string we would be changing the sign of the two charges to accommodate this configuration into that of Fig.~\ref {fig:defects} bottom right, and also increasing the magnetization in the direction of a [111] field.}\BibitemShut {Stop}%
\end{thebibliography}%
\end{document}